# Using Search Engine Technology to Improve Library Catalogs

Dirk Lewandowski

Fakultät Design Medien Information, Department Information
Hamburg University of Applied Sciences
Hamburg, DE

This is a preprint of an article accepted for publication in

*Woodswort, Anne (ed.): Advances in Librarianship 32(2010)*

# Using Search Engine Technology to Improve Library Catalogs

Abstract

This chapter outlines how search engine technology can be used in online public access library catalogs (OPACs) to help improve users' experiences, to identify users' intentions, and to indicate how it can be applied in the library context, along with how sophisticated ranking criteria can be applied to the online library catalog. A review of the literature and current OPAC developments form the basis of recommendations on how to improve OPACs. Findings were that the major shortcomings of current OPACs are that they are not sufficiently user-centered and that their results presentations lack sophistication. Further, these shortcomings are not addressed in current 2.0 developments. It is argued that OPAC development should be made search-centered before additional features are applied. While the recommendations on ranking functionality and the use of user intentions are only conceptual and not yet applied to a library catalogue, practitioners will find recommendations for developing better OPACs in this chapter. In short, readers will find a systematic view on how the search engines' strengths can be applied to improving libraries' online catalogs.

Keywords: Search engines, online catalogs, ranking, information seeking behavior, query types

## Introduction

For some years now libraries' online public access catalogs (OPACs) have been competing with, if not threatened by, Web search engines. Although it has not yet been agreed which way this threat should be answered – it is certain that search engines will now remain a rival of libraries and their catalogs. As a consequence, some libraries have taken advantage of search engines by outfitting their OPACs with this new technology in order to provide and maintain quicker access to desired content.

This technical improvement, however, is only one side of the coin. It should also be mentioned that despite the technological advantage and the improved content in the library catalogs, there is no guarantee that users will acknowledge that the OPAC as their primary or, indeed, only instrument for their search for academic content. In order to achieve this, it would be necessary not only to train the users sporadically, but to establish systematic information literacy programs. Although it is not the focus of this chapter, it is nevertheless an important issue to mention at the outset in order to make clear that there cannot be only a technical solution for the "OPAC problem".

Now, what can be done from a technical standpoint? In order to identify any technical requirements, it is necessary to examine the user. Apart from the OPAC, where else does academic research take place? General Web search engines rank first (Google being the most widely used). Others include academic search engines, interdisciplinary databases, professional databases, academically-oriented social software as well as the listings of publishers and online-booksellers. It is clear that the OPAC is merely one alternative among many and that if the OPAC is to remain relevant for more than mere stock retrieval of individual titles, it will clearly have to take position beside its rivals in the future.

During the process of digitalization and in the case of "digital only," definition of library material has frequently been requested. Libraries, which until now have defined themselves by their physical stock, are experiencing problems defining exactly what belongs to their collections. Should only printed stock be included? Should they include academic content that is freely available on the Web or only the licensed databases? Here, libraries are primarily asked which search path the users wish to follow. Do they want to begin searching in the local materials and then expand to additional collections when required? Would they prefer a "top-down approach" whereby the users initiates their search in a library before being guided to the locally available stock in a subsequent step? The answer to this is crucial: it means we are either dealing with an OPAC approach or with that of a academic search engine, which would naturally significantly influence the assembly of the system.

In this chapter, I will compare the OPACs system to some well known internet search engines. It is especially important to take Web search engines into consideration because they define the standards upon which other information systems (i.e. not only the library-search systems) will have to act in order to remain accepted by the users.

On the one hand it can be said that, due to their assembly and by responding to the user's characteristics, search engines educate their users towards a "bad" user's attitude. On the other hand, the search engines have shown that even simple requests can be answered satisfactorily by means of elaborate ranking systems. In this respect, search engines are role models because they cater to the actual search behavior of their users (i.e. their research knowledge) in an attempt to optimize the results.

The influence of Web search engines on the users' search behavior should not be underestimated. Nevertheless, there is another reason why they should be compared to library-based methods: Web search engines such as Google have developed services aimed at the core of the libraries, e.g., Google Books and Google Scholar. These services use search-engine technology and elaborate ranking systems for searches in "library contents." Specialized search engines, however, are not the only source of inspiration for library services. An examination of the ranking of general Web search engines is recommended.

The next section will begin with an overview on the stage of development of modern OPACs. It provides a description of the existing deficiencies of current OPACs and how far they can be improved in terms of a search-engine orientation. It will be argued that OPACs can only become a competitive alternative when they can impress with a mature and user-friendly ranking. Furthermore, it will be shown which factors can be used for such a ranking. Finally, suggestions for future OPAC advancement in research and practice will be made.

## OPAC's current state of development

In this chapter it will be assumed that the OPAC provide the central access to libraries' contents. I will then deal with the domains' spectrum of content, catalog enrichment (and user participation), and discovery.

As a general rule, the spectrum of content of a library is not fully represented in its catalogs (Lewandowski, 2006). This is due to the prevalent lack of journal articles, articles from anthologies, and contents of the library-licensed databases within the libraries' catalogs. Newer OPACs tackle this problem by adding further titles (articles) and an automatic search expansion to additional databases. The user remains unaware however, of the collections he is searching and the amount of information covered. For instance, it is not clear to users during a regular query which journal articles are covered. Are they effectively presented with the articles of all existing journals within the library? With the articles of all library-licensed electronic journals? Or is it an entirely self-contained collection independent from the library, which neither covers the collection completely nor is limited to it? While the idea of adding results from further sources to a search is surely a good idea, we can see that it is executed insufficiently. What would be needed is a systematic expansion of the library catalog with regard to the library’s holdings of articles.

With regard to external databases, the result is similar: here, there is an attempt to integrate further data sources by means of a federated search and, more recently, the establishment of "complete indices." The first case, however, only interrogates a limited number of databases and accounts only for a strongly limited number of results per database. The second case tries to avoid these issues as well as the performance problems that accompany the federated search. This concerns a seminal approach although the index assembly can only be achieved with great effort, which is not the least due to problems dealing with licensing rights.

When it comes to the catalog enrichment, two sources can be generated that support the admission of titles: On the one hand, additional information is selectively purchased, or complemented during the process of compiling catalog data; and on the other hand, library users themselves are urged to generate additional "user-generated content" for the titles. While the first case mainly deals with indexes, blurbs and possible reviews, it is the users who must amend ratings and reviews. As is known from all systems dealing with *user-generated contents,* the crucial factor remains the achievement of a critical group of users who are actually willing to contribute to the content. It has to be said here that only a fraction of those users wishing for user-generated content agree on generating it themselves. Taking this into consideration, it is strongly recommended that data be exchanged between a preferably large number of libraries instead of relying on their individual and limited users.

The expression "discovery" signifies an exploration of databases which integrate both searching and browsing approaches. While the user is searching for information it is not yet clear if he is specifically *searching* or merely browsing through the collection. The distinction is questionable in many cases. Here, the act of searching must be seen as alternating searching and browsing. This means that users at some points enter search terms while at other times they sift through the set of results, influenced by the system. This phenomenon is known as an “exploratory search.”

Search systems offer so-called "drill-down menus," which help the user explore the set of results. For instance, results are refined by media type, key word, year of publication and so forth. . This provides the user with a simple means to limit the number of hits from the initial query in order to achieve a manageable number of results. Furthermore, a great advantage is provided by the suggestions offered in the drill-down menus generated from the initial pool of results, which means they constitute a dynamic reaction on the original set of results, as opposed to static browsing which draws upon pre-determined classifications and tables of contents.

Additional information from the enhanced catalog supports the user in validating the received results. Furthermore, enhanced descriptions of the results aid user’s evaluations, reducing the need to evaluate entire texts. When these two functions are connected with a list/shopping cart, they facilitate the explorative search for literature and display a clearly added value compared to sole searching and browsing approaches.

An overview of modern OPAC-developments in Europe indicate that they (Community Walk, 2011) generally support the users well when it comes to the refinement of results sets as well as the screening of the received results. However most of their problem lie with incomplete support of the target-oriented search and therein predominantly with ranking of the given results. Although the more recent OPAC-approaches orientate themselves towards the users, they stay bound to the traditional idea of information search through an information professional particularly in one regard: They assume the user to be capable of (and to acknowledge the necessity of) constricting the results set in order to receive a manageable number of results which are then fully screened. However, Web-search engine-trained behavior shows that users tend to rely strongly on the order of the results produced by the search engine, instead of implementing a further refinement of the results set themselves. Studies on selective behavior within the results set show a strong focus on the results that rank first. Another influential factor, next to position, is any emphasis within the result description (see below for more information on selective behavior). Users demand quick access to the results and are not willing to think about formulating the request for long. The initially generated results list becomes important in terms of it being the basis for the decision of if, and how, the search is to be continued. This makes it essential to design the initial results list in such a way that the first positions already display relevant results. Moreover, for a significant number of requests the results list is sufficient in order to find answers.

Considering the pros and cons of the OPAC-searches compared with Web search engines (Table 1), it can be seen that the strengths of the OPACs lie in the areas relevant for elaborate research by information professionals, while the search engines are strong in all areas related to broadly untrained users. Accordingly, OPACs offer a wide number of functions that can be used for the specific query but also require advanced knowledge of refinement techniques and search languages. Search engines, however, contain only a very limited number of functions for a broader search. Some of the functions are even restricted in terms of their operational reliability (Lewandowski, 2004; Lewandowski, 2008a). The second strength of the OPAC lies in the existence of metadata in the database that can be utilized during the search. Nevertheless, this metadata (as shown above) is used to support browsing rather than for specific research. Here, a real opportunity for improving the systems can be seen.

TAKE IN TABLE 1

## Query types and search intentions

The evaluation of search systems should always be oriented towards the queries that are put to this special type of search system. In order to test and optimize an individual system, it is helpful to use actual queries for an evaluation. Furthermore, it is essential to use different query-types in tests in order to cover the different search intentions of the users. The analysis of those queries in terms of the (potential) search intention, and optimization of the system towards these intentions can be seen as the key to successful responses to queries.

In information science, a differentiation is made between a *Concrete Information Need* (CIN) and a *Problem-Oriented Information Need* (POIN) (Frants V. I.,Shapiro, J. & Voiskunskii, V. G., 1997,). CIN asks for factual information, and is satisfied with only one factum. In the case of document retrieval, this means it is satisfied with only one document containing the required factum. In contrast, POIN requires a smaller or larger number of documents for satisfaction. Table 2 sums up the distinctive features of CIN and POIN.

When applying the problem-oriented and the concrete information needs to the search with OPAC, one can distinguish between a thematic search on the one hand and an item-specific search on the other. The second case is also known as *known-item search* (Kantor, 1976) because it is already known that the corresponding title *exists*. It remains merely to be found in the system.
TAKE IN TABLE 2

Broder (2002) differentiates between three different types of intentions when querying Web search engines: informational, navigational, and transactional. Navigational queries aim at a Web page that is already known to the user or which he assumes exists, for example homepages of companies (e.g., DaimlerChrysler) or people (e.g., John von Neumann). Such queries normally terminate in one correct result. The information need is satisfied when the requested page is found.

In contrast, informational queries require more than one document (POIN). The user wishes to be informed about a topic and therefore reads several documents. Informational queries aim at static documents to acquire the desired information, which makes further interaction on the Web page unnecessary.

Transactional queries, however, aim at Web pages offering the possibility for a subsequent transaction, such as purchasing a product, downloading data or searching a database.

Broder's differentiation is also applicable to the OPACs. Here too, exist different search intentions that have to be responded to satisfactorily by the same information system. While the navigational query equals the known-item search, and the informational query corresponds to the topic search, the transactional query correlates with the search for an adequate source in order to do further research (see Table 3). Today's OPACs are not attuned to this diversity of queries and the types of queries are not being discerned during evaluation, which differs from the evaluation of search engines (Lewandowski, in press; Lewandowski & Höchstötter, 2007). In the course of a more user-oriented approach, the future development of the OPACs should be conducted with regard to the different types of queries. Furthermore, the differentiation between the query types is essential for an appropriate ranking; knowing the user's intention is of tremendous importance for the accentuation of adequate documents. In other words: A successful ranking is impossible without acknowledging the user's intention!

TAKE IN TABLE 3

This raises the question as to how far we can gain access to the user's intention and the query types. Generally, libraries carry out user-specific studies such as surveys and smaller laboratory investigations (e.g., qualitative opinion polls, search tasks under observation, focus groups). In my opinion, these methods are barely adequate to produce the required data. Even more so, it is necessary to investigate and continuously monitor internal log files. Indeed, log-file investigations for library catalogs have taken place in the past (e.g. Hennies & Dressler, 2006; Lown & Hemminger, 2009; Obermeier, 1999; Remus, 2002). However, they have concentrated primarily on the lengths of queries, the usage of field search and amplified search functions or, in the case of Lown & Hemminger, on the usage of drill-down menus. The analysis of the queries was not a priority in these investigations.

## Ranking systems as a central means for information search

This section will deal circumstantially with the search and selection behavior of users when they employing Web search engines. Again, it is assumed that the search behavior of users is applied to other types of information systems and that these systems then need to adjust to the given behavior, rather than requesting a too demanding adjustment to the respective system. Afterwards, the typical ranking factors applied to Web search engines will be discussed and the efficiency for library OPACs will be questioned. I will suggest a set of suitable ranking factors as well as point out a general problem of ranking: The repeated bias towards the same results. For this problem I will also suggest a solution based on Web search engines.

The user behavior towards Web search engines can be characterized as follows:

- Queries in most cases contain only a few words with the majority consists of one word, followed by two-word queries. The average length of German queries is 1,7 words (Höchstötter and Koch, 2008), while English queries are longer due to specifics of individual languages. A shift in the user behavior towards longer queries cannot be detected.
- Studies have shown that the user behavior concerning query formulation and length does not differ between library catalogs and Web search engines (Hennies and Dressler, 2006). While Web search engines admittedly are adjusted to this query behavior to a L.,

M.T., O'Brien, M., & Smyth, B. M.T., O'Brien, M., & Smyth, B. J., Keenoy, Kevin., Levene, M.,& Yaari, E. Haridasan, M., Brynjarsdóttir, H., Xia, L ; Joachims, T., Gay, G., Granka, L., Pellacini, F., Pan, B., great extent, conventional OPACs display very long results lists that are only sorted according to the age of the data.

- Selection behavior within the search-engine results lists show explicitly how much users rely on the engine-based ranking (Granka , Joachims, & Gay, 2004; Joachims, Granka, Pan, Hembrooke, & Gay, 2005; Loriga Haridasan, Brynjarsdóttir, Xia, Joachims, Gay, Granka, Pellacini, & Pan, 2008). Not only do a significant number of users limit themselves to screening the first page of results, (Höchstötter & Koch, 2008) they also focus heavily on the top results.

  Despite the fact that not only the position of the result is crucial but also the description of the result within the results list (Lewandowski, 2008b). Studies during which the order of the result sets were manipulated, have shown that the presentation of low-ranking results does not have a great impact on the selection behavior (Bar-Ilan, Keenoy, Levene, & Yaari, 2009; Keane, O'Brien, & Smyth, 2008).

  The characterization of the most important aspects of user behavior towards Web search engines as well as the related expectations towards other information systems, show the importance of an adequate ranking within the results list. This is true not only for success in the sense of efficiency and effectiveness of modern information systems, but also for their acceptance by the user. Commercial providers of search systems have known this for years and have adjusted their information systems to these conditions. Examples are *Google Scholar* and Elsevier's *Scirus.* Both systems administer very large databases and offer the user intelligently sorted results lists without neglecting possibilities of a complex research. In this respect, academic search engines can be seen as a role model for search applications in libraries (Lewandowski, 2006).

Also, regarding the factors used in the ranking, Web search engines can act as role models for other information systems. Although it is true that the insight gained from the ranking of Web contents is not applicable one-on-one to other content, the broad preliminary stages from this context can nevertheless help to improve the rankings in other contexts.

## Applying ranking to library materials

Already at an early stage, an attempt was made to apply ranking to a library-based inventory of titles. Buckland, Norgard, & Plaunt (1993) state that in online catalog, “the computer could be programmed to provide any one or combination of a variety of orderings" (p. 313). Nevertheless, most of the OPACs which rank the results are still limited to *text matching* and field weighting today (Dellit & Boston, 2007). Some library catalogs go beyond this and experiment with, for example, popularity factors (Flimm, 2007) as well as copy data and lending data (Mercun &Zumer, 2008; Sadeh, 2007). All of these experiments tried to integrate individual factors without systematically verifying adequacy and practical use.

Meanwhile, the core of the ranking problem has evolved from merely matching queries and documents (i.e. *text matching*), to quality evaluation of the potentially relevant documents gained through *text matching*. Considering the Web context, this can be explained by the sheer mass of documents that respond to a typical query on the one hand, and the very limited quality evaluation in the course of indexing on the other hand. Web search engines already try to discard so-called SPAM-documents and duplicates during this process. This cannot, however be compared to a quality-evaluation process, through the selection of a title, as implemented in libraries.

In the field of search engines, three sections of quality evaluation have developed which can serve as evidence for the improvement of ranking within library catalogs (Lewandowski, 2009):

- Popularity: The popularity of a document is referenced for its quality evaluation. For example, the number of user accesses and the dwell-time on the document is measured, as well as the linking of a document within its Web graph which is decisive for the ranking of Web documents. For this purpose, not only the number of clicks and links respectively is crucial, but weighted models are also implemented that enable a differentiated evaluation. These models are well-documented in literature (Culliss, G.A. 2003) DeanJ. A., Gomes, B.,Bharat, K., Harik, G., & Henzinger, M.H., 2002; Kleinberg, 1999; Page, L., Brin S., Motwani, R., & Winograd, T., (1999) and their main elements are applicable to document evaluation in library catalogs.
- Freshness: The evaluation of freshness is important for Web search engines in two respects: Firstly, it is a matter of finding the actual or rather relative publication and refresh-dates (Acharya A., Cutts, M.,Dean, J., Haahr, P., Henzinger, M., Hoelzle, U., Lawrence, S., Pfleger, K., Sercinoglu, O., & Tong, S., 2005). Secondly, the question is in which cases is it useful to display fresh documents preferentially. While the first point is omitted with regard to library content, the second point is highly relevant when it comes to different professional cultures. Whereas fresh literature would be favored in quickly changing disciplines such as the sciences, such a preference cannot be useful in historically oriented disciplines such as history and philosophy. Therefore the use of freshness should be limited.
- Locality: Although essential for search engines, evaluating documents according to their

proximity to the user has not often been taken into consideration in the library context ( Lewandowski, 2009). Proximity can be seen here as the physical location of the user such as in the library, on campus, or at home., as well as the physical location of the item such as a central library, branch library, the item's availability or unavailability, and total lack of a physical location in case of items that are available online.

Concerning library contents, a strong quality evaluation takes place due to the selection of the items by the library. However, a quality-oriented ranking is essential when it comes to responding to queries that are strongly oriented towards the *precision* of the results like a user who is searching for relevant titles in order to collect basic information on a certain field.

## Misunderstandings concerning ranking systems

The considerations reviewed so far show that quality as the aim of ranking has only been defined by means of auxiliary constructions such as weighted popularity. This might be lacking on the theoretical level, but as a pragmatic approach, it is a sustainable way of evaluating quality. It should be taken into consideration that a ranking system never changes the total quantity of results, but merely gives them a certain order. Therefore a ranking system provides an *additional* benefit compared to previously existing systems and does not limit the possibilities in any way for professional users in particular.

Unfortunately there exist some misunderstandings concerning ranking systems and not only in libraries. For instance, the argument is brought forward that one sole, clear and understandable sorting criterion is better than an elaborate ranking. It can be countered that, without taking into consideration the user's appraisal of either system, further sorting possibilities, in addition to relevance ranking can be offered without a problem.

Another misunderstanding is the idea that hitherto, OPACs work without a ranking. This assumption appears correct only at first sight. The question arises as to how far one should speak of a ranking system which sorts according to the year of publication. When ranking is considered simply as a non-random order of results though, sorting according to the publication date must also be seen as a form of ranking. In this case, one has to ask whether this form of ranking is the best solution for the user.

A third misunderstanding consists of the neglect of relevance ranking because it does not work well. Naturally it is hard to define "relevance" dependant on context. Nevertheless, this cannot obscure the fact that ranking according to *assumed* relevance is at least capable of offering a satisfactory order of results. The criterion for evaluation of such generated results lists of course can only be identified through the user.

The last misunderstanding is based on the exact opposite opinion to the third: Ranking is seen as solvable in an artless way. As a rule, standard processes of text matching by means of Term Frequency/ Inverted Document Frequency (TF/ IDF) are implemented here. This cannot lead to satisfactory results.

## Factors appropriate to rank library materials

I have shown that Web search engines go far beyond the limits a solely text-based ranking, and that such a ranking is not at all likely to be successful. Library catalogs and their scant bibliographic data present us with a similar problem. The text-based ranking needs to be amended with appropriate factors that are aimed at particular qualities of the documents. Below are demonstrated the relevant factors that can be employed. Focus is on four groups of ranking factors: text statistic, popularity, freshness and locality. Subsequent sections will refer to OPAC-specific ranking factors that cannot be summarized in one of these sections. Table 4 summarizes all ranking factors as mentioned in the following text.

TAKE IN TABLE 4

### Text statistic

Text-statistic systems normally use standard processes such as TF/IDF (Term Frequency/ Inverted Document Frequency). They can be successfully implemented within collections of text that have already gone through quality control when being accepted into a database (e.g., newspaper-databases). Here, library catalogs have a problem in that the items are generally too short to enable a successful ranking in accordance with a text-statistic system. The sole employment of text-statistic systems, however, leads to unsatisfactory results. Unfortunately, ranking is often likened to text-statistic ranking and thus it is assumed that the latter is generally not qualified for OPACs (e.g., in Beall, 2008).

Partly enriched catalog data results in the problem of a highly diverse range of items and this makes ranking by means of the same system impossible. Even more so, a diagnosis of additional information is necessary in advance. The documents are thus only consolidated into a concerted ranking after the initial ranking has taken place. Apart from the analysis of text, the sole existence of text can be regarded as a ranking factor. This way, items possessing a full text or at least a table of contents can be chosen opposite other items.

### Popularity

Popularity ranking can also be applied to library materials. Popularity could be ranked either on the basis of individual items or on the basis of a group of items. For instance, a group can be formed from all items by the same author, all items from the same publisher or all items within a book series.

All this popularity data is query-independent. Therefore, the values of an item do not have to be calculated at the time of the query but can be added to the item in advance. These values have to be refreshed only at certain intervals. Even if user ratings are taken into account, popularity measurements only need to be updated periodically even if they have changed over time.

Freshness

While freshness (measured by the year of publication) is the most-used ranking criterion in catalogs today, there is more to freshness than simply ordering results by date. It is hard to know when fresh items are a particular priority, as the need for freshness may differ from one discipline to another. For example, fresh items may be crucial to a computer-science researcher, but it may be a good idea to rely more heavily on authority than on freshness for ranking items related to philosophy.
It is therefore important to determine the need for fresh items and relate them to user needs?. The need for fresh items can be determined from the circulation rate for items from a certain group. Such groups can be a broad discipline or even a specific subject heading. Again, the "need-for-freshness“ factor can be calculated in advance and therefore does not take up calculation time when generating a results list according to a query.

Locality

Locality is a ranking factor that can take into account the physical location of the user as well as the availability of items in the results list. An item available at a local branch of the library could be ranked higher than items that are available only at another branch. One can also use lending data to rank items. For some users, items not currently available for lending may be of little or no use and could therefore be ranked lower.

The physical location of the user can also be used in ranking. When a user is at home, we can assume that they will prefer to find electronic items that can be downloaded (Mercun & Zumer, 2008). When they are at the library, this restriction will not apply, and items available in print form can be ranked alongside electronic results. The location of the user can be determined through the IP address of his or her computer.

Other relevant ranking factors

While the ranking factors mentioned above are adaptions of factors used by Web search engines, there are still some ranking factors that do not have a counterpart in one of the above named areas and only play a minor role. For instance, the size or type of the item may be considered. Monographs may be favored over edited books, books over journal articles (or vice versa), and physical materials over online materials. Moreover, the different needs of the individual subjects have to be taken into account. The exemplary comparison of informatics and philosophy shows that while informatics searches may rather be based on fresh literature from conference papers, philosophy may prefer monographs. Freshness data can be derived from lending data for certain subjects.

User groups may also determine ranking. For example, the needs of professors may differ greatly from the needs of undergraduate students. . Professors may need exhaustive searches for their research, whereas textbooks might be preferred in student searches, for instance.

Dividing library users into groups leads us to the question of personalization of result rankings. This requires individual usage data as well as click-stream data from navigation. However, collecting individual user data is always problematic and should be restricted to scenarios where the user knows what data is collected and has chosen this option. There are many possibilities for improving ranking when anonymous statistical data can be gathered from general or specific group behavior. Since this can be used, there is no real need to use individual user data.

The listed ranking factors are suitable to greatly improve ranking in library catalogs and indeed to implement any elaborate ranking at all. A compilation of ranking factors suitable for library materials is one thing, but only a good combination of ranking factors can lead to good results. Decisions concerning a combination depend heavily on individual collection and use cases.

## Arrangement of results lists

However, every ranking system raises certain problems that need to be solved. One of these problems is the bias towards the same results due to ranking algorithms. Due to the fact that the same formula is applied to every item, items with the same or a very similar ranking value will be found in neighboring positions on the list. If we assume an item that does not differ from a counterpart by any factor (e.g., circulation rate, locality), then these two items will have equal ranking. This could be counteracted by detecting and clustering very similar items. *Google Scholar*, for instance, clusters different versions of an article (e.g. publisher's versions and preprints) in its results lists. Of course it is still possible to gain access to the different versions, if needed. Such clustering of related items might seem simple at first sight but it turns out to be difficult to implement. Ranking however, requires clustering in order to secure satisfactory results.

Furthermore, ranking has to be supported by a deliberate rearrangement of the results list. The problem of similar items may be solved, but it is nevertheless possible that certain documents are rated higher than others. It is essential to think about this problem in advance. Not only should the user intention be taken into consideration for the query but it should also matter for the results, to wit, which result does the user expect according to the query.

It is crucial for the ranking system to detect whether the user asks a general or a specific question. For a universal query, it would be helpful to present a dictionary, a textbook, a relevant database, a corresponding journal and a fresh subject-matter related work. Thus, a small selection would be made, containing most likely at least one helpful item for the user. This example shows that it is necessary to think not only about suitable ranking factors but also about a suitable mixture of the results lists. None of these factors can be successful without the others.

## Conclusion

The main assumption of this chapter is that current OPACs (also those known as "Next-generation OPACs") provide support for research but still lack user orientation. This is found particularly in the processing of queries and the expedient generation of results lists. Appropriate ranking factors for the library context have therefore been identified and discussed.

I would like to conclude with some recommendations on how far library and information scientists, developers in business companies as well as responsible librarians and libraries can help to advance OPACS and library search engines.

1. In order to improve the OPACs it is essential to know exactly what the user's intention is. Here, systematic analysis of mass data from the OPAC-logfiles. Knowledge of the user's desire of information is necessary to enable improvement of the system. The logfiles alone can give information on the actual search frequency. These considerations could be the basis for library teams to think about how queries could ideally be responded to.
2. It is necessary to draw unsuccessful queries from the logfile. In addition, the click-paths of users having posted such queries should be followed. This should help to develop strategies to avoid zero-results and how to deal with those.
3. It could be helpful to analyze the clickthrough-data of the logfiles in order to identify different types of queries. The click frequency can give information on the type of query (Joachims, 2002; von Mach & Otte, 2009).
4. Development and implementation of suitable ranking systems is needed. A clear idea of the

assembly of "ideal results lists" should be developed before ranking systems are implemented. However, the traditional approach, based on the text- or field-based factors and that aims at a suitable ranking due to its weighting does not seem promising.

When looking at the OPACs today, we can see that they have fallen behind the modern developments as represented by the most-advanced search systems, namely Web search engines. These are in continuous development, and will continue to be the role model for other information systems. Libraries, as well as library catalog vendors are well advised to monitor search engine developments closely and analyze which of these developments could be adapted to improve the library catalogs.

TABLES

Table 1: Comparison of the strengths and weaknesses of OPACs and search engines

| | OPAC | Search engine |
|---|---|---|
| Simple searches | Weak, order of results sorted by date | Strong because of good ranking |
| Expanded searches | Range of functions | Marginal number of functions, faulty functions (!) |
| Order of results/ Ranking | Bad, sorted only by date | Good due to mature ranking and diversity within results lists |
| Presentation of results | Sparsely flexible due to author/title/year | Result description with static and context-related elements |
| Collection | Only part of the library offering | Integration of all collections provided by the search engine |
| Metadata | High-end quality data | Metadata is barely used; no own production |

Table 2: Concrete Information Need vs. Problem-Oriented Information Need (translated from Stock, 2007, p. 52)

| CIN | POIN |
|---|---|
| Thematic boarders are clearly defined. | Thematic boarders are not clearly definable. |
| It is possible to express the formulation of the query in exact terms. | The formulation of the query allows a variety of terms. |
| One fact-information is usually enough to cover the requirements. | Usually a diversity of documents must be found. It remains open as to whether the information need is covered. |
| The information problem is solved when the fact-information has been transmitted. | The transmission of literature information is possible to modify the information problem or generate a new need. |

Table 3: Applicability of query types according to Broder (2002) to library OPACs

| Query type according to Broder | Analogue query type in the OPAC | Example query | Explanation |
|---|---|---|---|
| Informational | Topic search | Collaborative tagging | Search for literature towards a certain topic, requesting a variety of |
| | | | documents |
| Navigational | Known-item search | Wolfgang Stock Information Retrieval | Search for evidence of a certain title; only one title is accepted |
| Transactional | Search for sources | Database of Court decisions | Search for a source/ database in which the research can be continued |

Table 4: Ranking factors for library catalogs (Lewandowski, 2009 modified)

| Group | Ranking factor | Note |
|---|---|---|
| Text statistic | Terms<br>- within bibliographic data<br>- within enriched data<br>- within full text | Bibliographic data does not contain enough text for a good term-based ranking.<br>The amount of text per catalog listing varies drastically, meaning that the same ranking algorithm cannot be applied to all three terms. |
| | Field weighting | Appearance of search term is weighted differently according to the field |
| | Availability of text<br>- review<br>- table of contents<br>- full text | The existence of additional information alone can lead to a better rating of an item |
| Popularity | Number of available local copies | Based on the individual item |
| | How often has the item been viewed? | Based on the individual item |
| | Circulation rate | Based on the individual item |
| | Number of downloads | Based on the individual item |
| | - author<br>- publisher<br>- book series<br>- user ratings<br>- citations | Based on either the individual item or a group of items |
| Freshness | Publication date | Based on the individual item (could also be measured by its relationship to a group of items to which it belongs, e.g. systematic group/ compartment) |
| | Accession date | Based on the individual item (could also be measured by its relationship to a group of items to which it belongs, e.g. systematic group/ compartment) |
| Locality | Physical location of the user (home, library, campus) | Location could be derived from IP address of a certain user |
| | Physical location of the item<br>- central library<br>- library branch<br>- electronically available (i.e. no physical location important to the user) | |
| | Availability of the item<br>- available as a download<br>- available at the library<br>- currently unavailable | |
| Other | Size of item (i.e. number of pages) | |
| | Document type (monograph, edited book, journal article etc.) | Could be related to the importance of certain document types within certain disciplines |
| | User group (e.g. professors, students, graduate students) | |